\documentclass[twocolumn,showpacs,aps,pra,amsmath,amssymb]{revtex4}
\usepackage{bm,color,bbm}
\usepackage{ulem} 
\usepackage{hyperref, mathtools,graphicx}

\newcommand{\beq}{\begin{equation}}
\newcommand{\eeq}{\end{equation}}
\newcommand{\bqa}{\begin{eqnarray}}
\newcommand{\eqa}{\end{eqnarray}}
\newcommand{\nn}{\nonumber}

\newcommand{\est}{\phi_{\rm est}}
\newcommand{\rmse}{{\rm RMSE}}

\definecolor{maroon}{rgb}{0.7,0,0}

\definecolor{ngreen}{rgb}{0.3,0.7,0.3}

\definecolor{golden}{rgb}{0.8,0.6,0.1}

\begin{document}
\title{Metrology with entangled coherent states - a quantum scaling paradox}
\author{Michael J. W. Hall}
\affiliation{Centre for Quantum Computation and Communication Technology (Australian Research Council), Centre for Quantum Dynamics, Griffith University, Brisbane, QLD 4111, Australia}

\begin{abstract}
There has been much interest in developing phase estimation schemes which beat the so-called Heisenberg limit, i.e., for which the phase resolution scales better than $1/n$, where $n$ is a measure of resources such as the average photon number or the number of atomic qubits.  In particular, a number of nonlinear schemes have been proposed for which the resolution appears to scale as $1/n^{k}$ or even $e^{-n}$, based on optimising the quantum Cramer-Rao bound.  Such schemes include the use of entangled coherent states.  However, it may be shown that the average root mean square errors of the proposed schemes (averaged over  any prior distribution of phase shifts), cannot beat the Heisenberg limit, and that simple estimation schemes based on entangled coherent states cannot scale better than $1/n^{1/4}$.  This paradox is related to the role of `bias' in Cramer-Rao bounds, and is only partially ameliorated via iterative implementations of the proposed schemes. The results are based on new information-theoretic bounds for the average information gain and error of any phase estimation scheme, and generalise to estimates of shifts generated by any operator having discrete eigenvalues.
\end{abstract}

\pacs{ 42.50.St, 03.67.-a, 06.20.Dk, 42.50.Dv}
\maketitle

\section{Introduction}

Phase estimation is ubiquitous in metrology, and forms the underlying principle for the estimation of physical quantities via interferometry. For example, a physical variable of interest, such as temperature or refractive index or pressure or gravitational wave strength, may influence the relative phase between two pathways in an optical interferometer, and estimation of the change in phase then allows estimation of the physical variable.

At the fundamental quantum level, the accurate estimation of (relative) phase requires being able to distinguish between members of a set of overlapping quantum states, where each member has undergone a different phase shift.  For optical phase shifts, the ability to do so very much depends on photon number properties.   For example, the overlap of two pure states $|\psi\rangle$ and $e^{-iN\phi}|\psi\rangle$, having a small relative phase shift $\phi\ll 2\pi$, is easily calculated to be 
\[ |\langle\psi|e^{-iN\phi}|\psi\rangle|^2 = 1-\phi^2 (\Delta N)^2 +O(\phi^4). \]
Hence, a correspondingly large photon number variance, on the order of $1/\phi^2$, is necessary for the overlap to be small enough to allow an accurate distinction between these states.  More generally, the maximum possible resolution of a given phase estimation scheme will depend on the resources available, such as the average photon number per probe state for optical probe states, or the number of atomic qubits of the probe state in Ramsey interferometry.

The optimal scaling of the resolution with the resources, for a given estimation scheme, is of considerable interest.  A common tool for determining such scalings is the quantum Cramer-Rao bound \cite{cr1,cr2}, which suggests, for example, that using the NOON state $[|n\rangle|0\rangle+|0\rangle|n\rangle]/\sqrt{2}$ as a probe state can achieve a phase resolution that scales as $1/n$ \cite{noon}, while using the $n$-qubit state $\otimes^n |z\rangle$ can achieve a phase accuracy scaling as $2^{-n}$ \cite{roy}.  The quantum Cramer-Rao bound similarly suggests that entangled coherent states can achieve a similar $1/n$ scaling as for NOON states \cite{jooprl, hirota}, and surpass NOON states for small values of $n$ \cite{jooprl, joopra}.

However,  recent results on the accuracy of phase estimation schemes, based on quantum information properties of the probe state, raise a scaling paradox \cite{prx}.  For example, they imply that a NOON state cannot achieve more than one bit of phase information per use, independently of $n$, and that the corresponding accuracy does not approach zero as $n$ is increased.  More generally, the results imply that the accuracy promised by the quantum Cramer-Rao bound, for a given probe state, often cannot  be achieved,  {\it unless} the phase shift is already known to about that accuracy! 

The scaling paradox arises essentially because the quantum Cramer-Rao bound has a restricted application, to regions in which the estimation scheme is locally unbiased.  For the above-mentioned phase estimation schemes, these regions are comparable in size to the accuracy promised by the Cramer-Rao bound.  This limitation can be overcome, to some extent, via considering iterative implementations of the proposed schemes.  Such implementations use multicomponent probe states, each component of which estimates a different bit of the phase shift.  In many cases (but not all), one can recover the promised scaling of the Cramer-Rao bound in this way, albeit with a large scaling factor in general \cite{prx}.

The new information-theoretic bounds for phase estimation yield general bounds on phase resolution and information gain, applicable to any phase estimation scheme, and take into account any prior information available about the phase shift.  They are reviewed below, and compared with the quantum Cramer-Rao bounds for the case of entangled coherent states in particular.  It is shown that using such a state as a probe state cannot achieve an average resolution that asymptotically scales better than $n^{-1/4}$, whereas, in contrast, {\it unentangled} coherent states can achieve a scaling of $n^{-1/2}$.  This is in contrast to the quantum Cramer-Rao bound scalings of $n^{-1}$ and $n^{-1/2}$ respectively.  Moreover, it is shown that it is not straighforward to achieve the latter $n^{-1}$ scaling for entangled coherent states (and may not even be possible to do so), due to the presence of a strong vacuum background contribution to the phase properties of such states. The behaviour of the two types of  bound for small average photon number $n$ is also examined.

\section{Phase estimation schemes}

As shown in Fig.~1, a generic phase estimation scheme involves application of a phase shift to a probe state, and subsequent estimation of the phase shift via some measurement. Such schemes can be very general, and include the case of complex (possibly adaptive) measurements made across the components of an entangled multicomponent probe state.  The generator $G$ of the phase shift will be taken throughout this paper to have integer eigenvalues, ensuring that the phase-shifted state $\rho_\phi$ satisfies the fundamental phase-shift property $\rho_{\phi+2\pi}=\rho_\phi$.  However, many of the results below hold for more general operators $G$ \cite{prx}.  

\begin{figure}[!ht]
\centering
\includegraphics[width=0.45\textwidth]{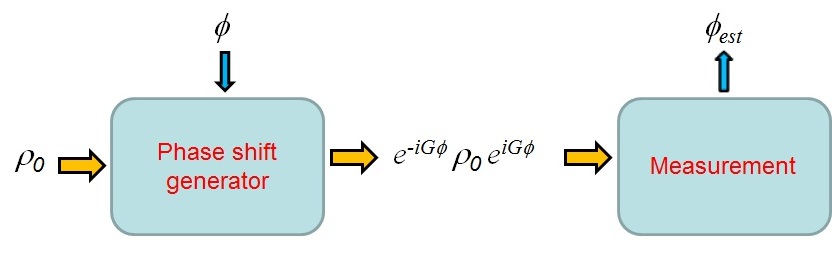}
\caption{Generic structure of phase estimation schemes.  A probe state $\rho_0$ undergoes a phase shift $\phi$, generated by some operator $G$, and a measurement on the probe state is used to make an estimate, $\phi_{\rm est}$, of $\phi$.  The phase shift may arise from the probe state passing through a particular environment (e.g., in one arm of an interferometer), or via some external modulation (e.g., in a communication scenario).  The probe state may comprise, for example, a single-mode optical field, a multimode field, an atomic qubit, or several such qubits.  The generator $G$ may be any suitable Hermitian operator defined on the Hilbert space of the probe state: for example $N$ or $N^2$ for a single-mode field having photon number operator $N$; $N_1+N_2+\dots+N_m$ for a multimode field comprising $m$ single-mode fields; or $\sigma^{(1)}_z+\sigma^{(2)}_z+\dots+\sigma^{(n)}_z$ for a probe state comprising $n$ atomic qubits.  The measurement may comprise, for example, individual measurements on components of the probe state (with some type of averaging of the individual outcomes to form an estimate), or a complex measurement across all components of an entangled probe state. }
\label{fig1}
\end{figure}

A central question of interest is how good can a given estimation scheme be?  The answer will in general not only depend on the probe state, the generator, and the measurement, but also on the measurement of performance used.  Several performance bounds are discussed in this section.

\section{Phase estimation bounds}

\subsection{Quantum Cramer-Rao bound}

One can first ask how well a given phase estimation scheme performs for a particular phase shift value $\phi$.  A natural measure of the performance for this case is the local root mean square error, $\rmse_\phi$, defined by
\beq \label{local}
{\rm RMSE}_\phi :=  \left[\int d\phi_{\rm est}\, p(\est|\phi)\,(\est-\phi)^2\right]^{1/2} , \eeq
where $p(\est|\phi)$ denotes the conditional probability of estimating $\est$ for an applied phase shift $\phi$.  Note that there is an ambiguity in defining the phase reference interval over which the integration is performed; however, the bounds given below are independent of the choice of this interval \cite{berrypra}.

The quantum Cramer-Rao bound is valid for the special case that the estimate is {\it locally unbiased} at $\phi$, i.e., when 
\[ \langle \est\rangle_{\phi'}:=\int d\phi_{\rm est}\, p(\est|\phi')\,\est=\phi' \]
for all $\phi'$ in some neighbourhood of $\phi$.  In particular, for locally unbiased estimates one has \cite{cr1,cr2}
\beq \label{cr}
\rmse_\phi \geq \frac{1}{\sqrt{F_\phi}} \geq \frac{1}{2\Delta G} ,\eeq
where $F_\phi$ denotes the Fisher information of the estimate, and $\Delta G$ denotes the root mean square spread of the generator for the probe state.  The second inequality is sometimes referred to as the Helstrom-Holevo bound \cite{hh}, and is saturated for pure states \cite{cr2}.

It is easy to show, for example, that if the probe state comprises a NOON state of a single-mode field, then Eq.~(\ref{cr}) reduces to $\rmse_\phi \geq 1/n$ \cite{noon}.  Note also that if the probe state comprises a tensor product of $m$ identical components, with $G=G_1+\dots G_m$, then $(\Delta G)^2=(\Delta G_1)^2 +\dots (\Delta G_m)^2 = m(\Delta G_1)^2$, and hence the righthand term scales as $1/\sqrt{m}$ \cite{cr2}.

\subsection{Local unbiasedness}

The requirement of local unbiasedness for the validity of Eq.~(\ref{cr}) is surprisingly strong.  Suppose, for example, that one in fact knows beforehand that the phase shift applied to the probe state is $\phi_0$.  Clearly, there is then no need for any physical measurement to make a perfect estimate: one simply takes $\est\equiv\phi_0$.  The local root mean square error in Eq.~(\ref{local}) then vanishes, i.e., $\rmse_{\phi_0}=0$.  At first sight this appears to contradict the quantum Cramer-Rao bound in Eq.~(2), since the latter implies that in the case the probe state is a NOON state, then  $\rmse_{\phi_0}\geq 1/n$.  There is, of course, no real contradiction: the `perfect estimate' is not locally unbiased, so that Eq.~(1) does not apply (one has $\langle \est\rangle_{\phi}=\phi_0$, rather than $\phi$).  However, this example shows that the restriction to locally unbiased estimates can give misleading indications as to what may be possible.  More practical examples will be given in later sections.

One may remove the requirement for local unbiasedness via a more general form of the quantum Cramer-Rao inequality due to Helstrom \cite{hel}:
\bqa \nn (\rmse_\phi)^2 &\geq& \left[\langle \est\rangle_\phi-\phi\right]^2 + \frac{[\partial \langle \est\rangle_\phi/\partial\phi]^2}{F_\phi}\\
&\geq&  \left[\langle \est\rangle_\phi-\phi\right]^2 + \frac{[\partial \langle \est\rangle_\phi/\partial\phi]^2}{4(\Delta G)^2} , \label{gen}
\eqa
where $\langle \est\rangle_\phi$ is defined above, and the second inequality follows via the corresponding inequality in Eq.~(\ref{cr}).  Note this formula implies that Eq.~(\ref{cr}) also holds in the case of estimates with a locally {\it constant} bias (i.e., with $\partial \langle \est\rangle_\phi/\partial\phi=1$). 

More significantly, Eq.~(\ref{gen}) implies, for example, that one {\it cannot} obtain a $1/n$ scaling of the local root mean square error for NOON states, if the estimate is biased at $\phi$.  Thus, only locally unbiased estimates tend to be considered when using Cramer-Rao bounds to compare the accuracies of various phase estimation schemes. This restriction, however, greatly limits the applicability of such bounds to many of the schemes proposed in the literature, which are sometimes only locally unbiased for a finite set of phase shift values.

Finally, another way remove the requirement for local unbiasedness is to replace the measure of performance by a different quantity, the {\it local precision}, defined by \cite{cr2}
\[ P_\phi:=\left[ \int d\phi_{\rm est}\, p(\est|\phi)\,\left(\frac{\est}{|\partial \langle \est\rangle_\phi/\partial\phi|}-\phi\right)^2 \right]^{1/2} . \]
This quantity satisfies the same inequalities as $\rmse_\phi$ in Eq.~(\ref{cr}), but is valid for {\it any} estimate, whether biased or unbiased \cite{cr2}.  In particular, the local precision scales as $1/n$ for NOON states, for all values of $\phi$ \cite{noon}.  However, the operational meaning of the local precision as a measure of performance is not clear, due to the nonlinear scaling term in the denominator.  The only exception appears to be if this term is a constant, $k$ say, over the range of of phase shifts of interest, as it is then natural to replace the estimate $\est$ by the rescaled estimate $\est'=\est/k$.  But for this case the local precision of $\est$ is just the local root mean square error of $\est'$, and the bound reduces to a particular case of the general Helstrom bound above.

\subsection{Information-theoretic bounds}

Recently, new bounds on phase estimation have been derived, which are valid for both biased and unbiased estimates, and which allow prior information about the phase shift (such as used to make the `perfect estimate' in the above example) to be taken into account.  The first bound limits the Shannon mutual information between the phase shift and the estimate, $H(\est|\phi)$, i.e., the amount of information which can be gained per estimate about the phase shift, and is given by \cite{prx}
\beq \label{inf}
H(\est|\phi) \leq A_G(\rho_0) \leq H(G|\rho_0) . \eeq
Here $A_G(\rho_0)$ is the increase in von Neumann entropy corresponding to a projective measurement of $G$ on the probe state, also known as the $G$-asymmetry of the probe state \cite{gasymm}, and $H(G|\rho_0)$ denotes the Shannon entropy of the probability distribution of $G$ for the probe state. 

Note, for example, that estimation using a NOON state in an interferometer, with $G=N_2$ where $N_1$ and $N_2$ are the number operators of the modes, it follows that $A_G(\rho_0)=H(G|\rho_0) = \log 2$.  Hence no more than one bit of information about the phase shift can be extracted from a NOON state, independently of $n$.  A similar result holds if $G$ is replaces by any (possibly nonlinear) function of $N_1$ and $N_2$ \cite{prx}.  Note, however, that this one bit can still be very useful, as part of a more complex `bit-by-bit' phase estimation scheme based on multicomponent probe states, as will be seen in the next section.

One may also derive an information-theoretic bound for the {\it average} estimation error, i.e., for the root mean square error 
\beq \label{av}
\rmse :=  \left[\int d\phi\,d\phi_{\rm est}\, p(\est, \phi)\,(\est-\phi)^2\right]^{1/2}  \eeq
of the estimate, where $p(\est, \phi)$  the joint probability distribution for $\est$ and $\phi$.  Note that if $\wp(\phi)$ denotes the prior probability density for the applied phase shift, then one has the relations $p(\est, \phi)=p(\est| \phi)\wp(\phi)$ and $(\rmse)^2 = \int d\phi\,\wp(\phi)\,(\rmse_\phi)^2$.  The latter relation shows that the average estimation error is a suitable measure of the average performance of the estimate, which takes prior information about the phase shift into account via $\wp(\phi)$.  

As shown in Ref.~\cite{prx}, Eq.~(\ref{inf}) implies the lower bounds
\bqa \nn
\rmse &\geq& (2\pi e)^{-1/2} e^{H(\phi)} e^{-A_G(\rho_0)}\\
&\geq& (2\pi e)^{-1/2} e^{H(\phi)} e^{-H(G|\rho_0)} , \label{avbound} \eqa
for the root mean square error, which strengthen and generalise earlier results in the literature \cite{nair, rapcomm}.  Here $H(\phi):=-\int d\phi \,\wp(\phi) \log \wp(\phi)$ denotes the Shannon entropy of the prior distribution, and reduces to $\log 2\pi$ in the case of random phase shifts, i.e., when $\wp(\phi)\equiv 1/2\pi$.  The second inequality is saturated for pure probe states.

For example, again taking a NOON state with $G=N_2$ one finds the lower bound $\rmse\geq (8\pi e)^{-1/2} e^{H(\phi)}$, for all values of $n$.  Hence, {\it no} scaling of the root mean square error with $n$ is possible in this case --- contrary to what is suggested by the corresponding quantum Cramer-Rao bound $\rmse_\phi\geq 1/n$ following from Eq.~(\ref{cr}).  This apparent scaling paradox arises because of the restriction of the latter bound to locally unbiased measurements, and is examined more closely in the following sections, and with particular reference to entangled coherent states in Sec.~V.  

Finally, note that, for the `perfect estimate' discussed at the beginning of Sec.~III~B, one has the prior distribution $\wp(\phi)=\delta(\phi-\phi_0)$.  Hence $H(\phi)=\infty$, and the lower bound in Eq.~(\ref{avbound}) yields $\rmse\geq 0$ --  consistent with the fact that $\rmse=0$ for this estimate.  This highlights the fact that the information-theoretic bounds, unlike the quantum Cramer-Rao bound in Eq.~(\ref{cr}), are valid whether or not the estimate is locally unbiased.

\section{Scaling paradox and iterative estimation schemes}

The quantum Cramer-Rao bound in Eq.~(\ref{cr}) suggests that, all else being equal, probe states should have as large a variance of $G$ as possible, whereas the information-theoretic bound in Eq.~(\ref{avbound}) suggests that they should have as large an entropy of $G$ as possible.  These can lead to conflicting results in various cases, as seen in the NOON state examples above, which have large variance but small entropy.  More generally, for generators with integer eigenvalues, if follows from Eq.~(\ref{avbound}) that \cite{prx}
\[ RMSE \geq \frac{1}{\sqrt{2\pi e \left[ (\Delta G)^2 +1/12\right]}} ,\] 
which scales as $1/\Delta G$ for $\Delta G\gg 1$.  While this is consistent with the quantum Cramer-Rao bound in Eq.~(\ref{cr}), the bound in Eq.~(\ref{avbound}) is typically much stronger.

 For example, for a single-mode field with average photon number $\langle N\rangle$, and a nonlinear phase shift generator $G=N^k$, 
one can easily find states with the scalings
\[ \Delta G \sim \langle N^{2k}\rangle^{1/2} \sim \langle N \rangle^k, \]
suggesting via the quantum Cramer-Rao bound that a local root mean square error scaling as $1/\langle N\rangle^k$ is possible \cite{luis,joopra} --- greatly improving on the Heisenberg scaling limit $1/\langle N\rangle$ for $k>1$.  However, noting that the entropy of $N$ and $N^k$ are identical for any state,  the information-theoretic bound may be used to show that the average estimation error is always bounded by \cite{prx}
\beq \rmse \geq \frac{e^{H(\phi)}}{\sqrt{2\pi e^3} \langle N+1\rangle} . \eeq
Hence, no better than Heisenberg scaling is possible for the average performance of the estimate, in such nonlinear scenarios. 

The paradoxical differences in scalings between the two bounds arise from the restriction of the quantum Cramer-Rao bound to locally unbiased measurements.  In particular, many of the proposed phase estimation schemes in the literature are only locally unbiased, or approximately so, for a small range of phase shift values.  Indeed, this range is often of the same order of magnitude as the lower bound in Eq.~(\ref{cr}), as is discussed in more detail in Refs.~\cite{prx} and \cite{berrypra}.  In such cases, the quantum Cramer-Rao bound can only be applied if the phase shift is known to fall within this narrow range, i.e., if the phase is already known to an accuracy comparable to that promised by the bound!  Thus, while these schemes appear to offer increased phase resolution with increasing resources $n$, they can only achieve this if the phase shift before measurement is correspondingly known more and more accurately with increasing $n$.

In contrast, the information-theoretic bound in Eq.~(\ref{avbound}) explicitly takes prior information about the phase shift before measurement into account, via the term $e^{H(\phi)}$.  For example, if the phase shift is known to be randomly taken from an interval of width $W$, then this term gives a scaling factor of $W$ for the $\rmse$.  Thus, the bound explicitly separates out the scaling contributions from the choice of probe state on the one hand, and prior information about the phase shift on the other.

\begin{figure}[!ht]
\centering
\includegraphics[width=0.45\textwidth]{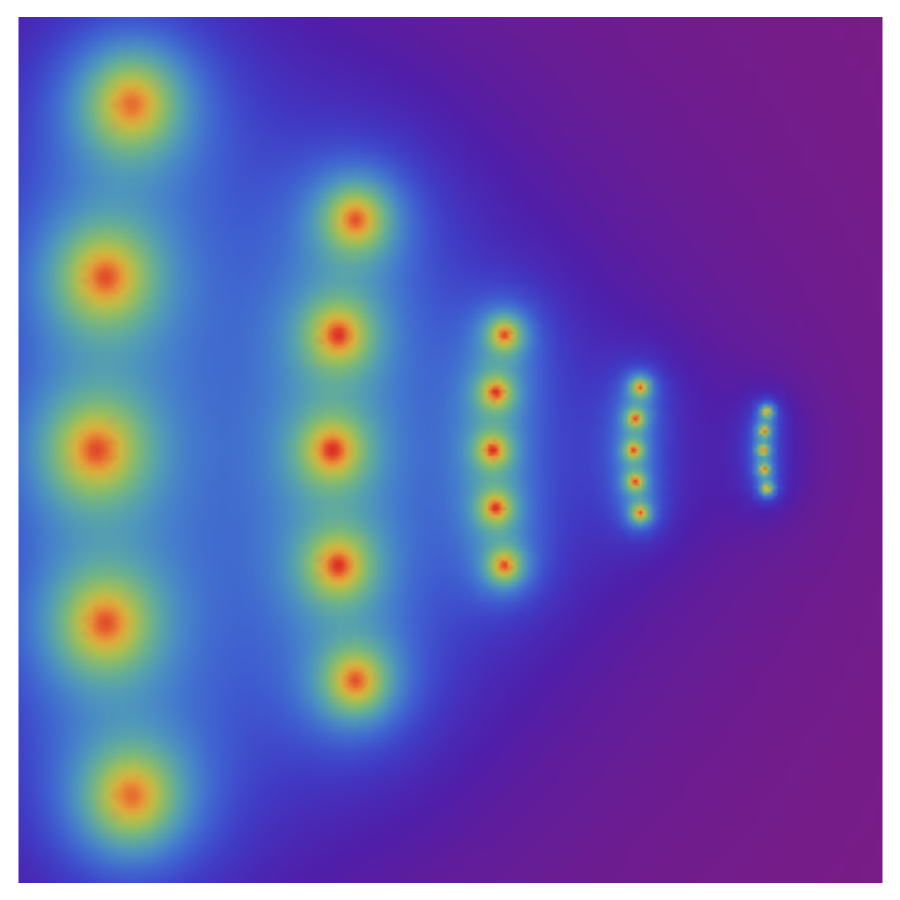}
\caption{Iterative phase estimation schemes. In this conceptual diagram, each circular region represents a component of a multicomponent probe state.  The components corresponding to the five largest regions, on the left of the figure, are used to estimate the first bit of $\phi/(2\pi)$; the next   five components are used to estimate the second bit, and so forth.  For example, each size region may correspond to a different NOON state, with $n=1, 2, 4, 8, 16$ from left to right, or to a different entangled coherent state, with exponentially increasing $\alpha$ from left to right.}
\label{fig2}
\end{figure}

Finally, it is of interest to note that the scaling promise suggested by quantum Cramer-Rao bounds can sometimes (but not always) be achieved via the implementation of {\it iterative} versions of the proposed schemes, based on multicomponent probe states.  The idea is that different components, each having a different number of resource $n$, are used to estimate successive bits of the phase shift, as depicted in Fig.~2.  When the total number of resources required for the probe state are added up (e.g., the total photon number or the total number of atomic qubits), and compared with the resolution of the estimate, one can often obtain the scaling suggested by the quantum Cramer-Rao bound for a single-component probe state, although generally with a larger scaling constant.  Several examples have been discussed elsewhere \cite{prx}, including a case where a $2^{-n}$ scaling, suggested via the quantum Cramer-Rao bound \cite{roy}, cannot be achieved even with an iterative implementation.

\section{Entangled coherent states}

Entangled coherent states (ECS) may be defined in various ways (see Ref.~\cite{sanders} for a recent review of such states), but for our purposes will be taken to be superpositions of tensor products of Glauber coherent states.  In particular,  consider an ECS of the form
\beq \label{ecs} |\psi_\alpha\rangle = \frac{1}{\sqrt{2(1+e^{-|\alpha|^2)}}} \left(|\alpha\rangle |0\rangle +|0\rangle|\alpha\rangle\right) . \eeq
Such two-mode states have been shown to share precisely 1 bit of entanglement \cite{hirbit}, and maximally violate a Bell inequality of the Clauser-Horne-Shimony-Holt type \cite{jeong}.  

The above ECS is seen to be similar in form to NOON states, with number states being replaced by coherent states, and indeed it has been shown \cite{jooprl,hirota, joopra} that for large $|\alpha|$ the quantum Cramer-Rao bound scales in the same way with the total average photon number, 
\[ n:=\langle N_1+N_2\rangle  ,\]
for both types of state \cite{jooprl,hirota, joopra}. Further, for small values of $n$, ECS outperform  NOON states, as quantified by the Cramer-Rao bound, both for linear and nonlinear phase shifts \cite{jooprl,joopra}.  This  suggests that ECS may be valuable states for quantum metrology.

However, in light of the preceding sections, it is seen that it is important to take into account that the quantum Cramer-Rao bound is restricted in operational significance, to the case of estimation schemes that are (at least approximately) locally unbiased over the phase shift range of interest.  Hence, the performance of the average estimation error for schemes based on entangled coherent states is carefully assessed below, using the more general information-theoretic bound in Eq.~(\ref{avbound}).

\subsection{Large average photon number}

Consider  
an inteferometric setup in which the second component of the ECS $|\psi_\alpha\rangle$ 
is subjected to a phase shift $\phi$ \cite{jooprl}.  For a linear phase shift the generator is then 
 $G=N_2$.  To evaluate the bound for the $\rmse$ in Eq.~(\ref{avbound}), one then 
needs to determine the entropy of $N_2$ for the probe state.  

Taking $\alpha>0$ without loss of generality, let $\tilde{p}_m:=e^{-\alpha^2}\alpha^{2m}/m!$ denote the photon number probability distribution for a single-mode coherent state.  It is straightforward to calculate from Eq.~(\ref{ecs}) that the probability distribution of  $N_2$ for the ECS $|\psi_\alpha\rangle$ is then
\[ p_m = \frac{\tilde{p}_m +\delta_{m0}(1+2\tilde{p}_0)}{2(1+e^{-\alpha^2})} \approx \frac{1}{2} \left( \tilde{p}_m + \delta_{m0}\right) \]
where the approximation is valid for sufficiently large $\alpha$.  

Thus, the distribution of $N_2$ is well approximated by an equal mixture of the Poisson distribution $\{\tilde{p}_m\}$ with the trivial distribution $\{\delta_{m0}\}$ -- where these distributions are effectively nonoverlapping since $\tilde{p}_0\ll 1$ for large $\alpha$. Hence, as is easily verified, the entropy of this distribution is approximately equal to the average of the entropies of the Poisson and trivial distributions, plus $\log 2$.  The above approximation for $p_m$ also implies that $n=2\langle N_2\rangle\approx \alpha^2$. Finally, since the Poisson distribution $\{\tilde{p}_m\}$ is well approximated by a Gaussian distribution of the same mean ($\alpha^2$) and variance ($\alpha^2$) for large $\alpha$, where the entropy of a Gaussian of variance $v$ is known to be $(1/2)\log 2\pi e v$, it follows that 
\[ H(G|\psi_\alpha) \approx \frac{1}{4} \log 2\pi e \alpha^2 +\log 2 \approx \frac{1}{4} \log 2\pi e n +\log 2 . \]
Substitution into Eq.~(\ref{avbound}) gives the approximate lower bound
\beq \label{ECS}
\rmse(ECS) \gtrsim \frac{e^{H(\phi)}}{2(2\pi e)^{3/4}n^{1/4}} \approx \frac{0.060\, e^{H(\phi)}}{n^{1/4}} \eeq
for phase estimation based on an ECS, in the limit of large $n$.  

This  scaling as $1/n^{1/4}$ strongly  contrasts with  the $1/n$ scaling of the quantum Cramer-Rao bound for ECS \cite{jooprl,hirota}.  Indeed, as seen in Fig.~ 3, the scaling in Eq.~(\ref{ECS}) is surpassed by the corresponding scaling for {\it unentangled} coherent probe states of the form $|\alpha\rangle|\alpha\rangle$, for which one finds the approximate lower bound
\beq \label{coh}
\rmse(COH) \gtrsim \frac{e^{H(\phi)}}{\pi{e}\sqrt{2n}}\approx \frac{0.083\,e^{H(\phi)}}{n^{1/2}} \eeq
in the limit of large $n$, via a similar calculation (a numerical check shows that both approximate bounds are accurate for $n\gtrsim 20$). However, both types of probe state improve on the scaling corresponding to use of a single NOON state, for which Eq.~(\ref{avbound}) gives
\beq \label{NOON} \rmse(NOON) \geq \frac{e^{H(\phi)}}{\sqrt{8\pi e}} \approx 0.121\,e^{H(\phi)} \eeq
for {\it any} (integer) value of $n$.  

\begin{figure}[!ht]
\centering
\includegraphics[width=0.45\textwidth]{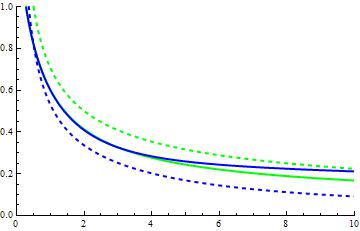}
\caption{Dependence of quantum Cramer-Rao and information-theoretic bounds, on the total average photon number $n$, for entangled coherent states and factorisable coherent states. The dotted curves show the quantum Cramer-Rao bound for the {\it local} performance, $\rmse_\phi$ in Eq.~(\ref{cr}), as a function of $n$, with the upper (green) dotted curve corresponding to probe states of the form $|\alpha\rangle|\alpha\rangle$, and the lower (blue) dotted curve to ECS probe states as per Eq.~(\ref{ecs}).  Thus the Cramer-Rao bound suggests that ECS have a better local performance.  However, the solid curves show the information-theoretic bound for the {\it average} performance, $\rmse$ in Eq.~(\ref{avbound}), with the upper (blue) solid curve corresponding to ECS probe states and the lower (green) solid curve corresponding to factorisable coherent states (for the choice of a random prior distribution, $\wp(\phi)=1/2\pi$).  Hence, it is seen that the bound on average performance is very similar for both types of state, for small values of $n$ ($n<5$), but is lower (i.e., better) for factorisable coherent states as $n$ increases -- as expected from the asymptotic scalings in Eqs.~(\ref{ECS}) and (\ref{coh}).}
\label{fig3}
\end{figure}

\subsection{Iterative implementations}

It is known that, despite Eq.~(\ref{NOON}), iterative implementations based on NOON-state components can recover the $1/n$ scaling suggested by the quantum Cramer-Rao bound, albeit with a larger scaling constant \cite{prx}.  It is also shown below that iterative implementations based on {\it factorisable} coherent states can achieve the $1/\sqrt{n}$ scaling 
 of  the corresponding quantum Cramer-Rao and information-theoretic bounds.  This is because the canonical phase distribution for the second mode of the state $|\alpha\rangle|\alpha e^{-i\phi}\rangle$ 
(this distribution 
is expected to be optimal in the case of an unknown phase shift on such a state \cite{hel74,rapcomm}) is approximately Gaussian, with mean $\phi$ and variance $1/4\alpha^2=1/2n$ \cite{hel74}, and hence can resolve phase to an accuracy on the order of $1/\sqrt{n}$, where typically $M$ repetitions will be required to do so with near certainty, for $M$ typically in the range 4-8 \cite{prx}.    However, as will be seen, it is not clear whether a similar conclusion holds for iterative implementations based on ECS.

In particular, consider first a multicomponent probe state $\rho_0$ comprising: $M$ copies of such factorisable states with total average photon number $n_1=1$, to estimate the first bit of $\phi/2\pi$; $M$ copies with $n_2=2^2$ to estimate the second bit; etc., culminating with $M$ copies with $n_m=2^{2(m-1)}$ to estimate the $m$th bit (see Fig.~2).  The total average photon number of this multicomponent probe state is then $n=M\sum_{j=0}^{m-1}2^{2j} = M(4^m-1)/3$.  Further,  the generator $G$ is the sum of the photon number operators over all components of $\rho_0$.  Since each component has an approximately Gaussian photon number distribution (with variances $n_1/2, n_2/2,\dots,n_m/2$), and the variance of the sum of independent Gaussian variables is a Gaussian with variance equal to the sum of the component variances, the entropy $H(G|\rho_0)$ can be approximated as 
\begin{eqnarray*} 
H(G|\rho_0) &\approx&  \frac{1}{2} \log \left[2\pi e M(n_1+\dots +n_m)/2\right]\\ &=& \frac{1}{2} \log [2\pi e n]. 
\end{eqnarray*}
Substitution into the information-theoretic bound in Eq.~(\ref{avbound}) then gives precisely the same bound as in Eq.~(\ref{coh}) above.  Since the iterative scheme has, by construction, a resolution of $\approx (2\pi)/2^{m+1}\approx \pi\sqrt{M/3n}$, it follows that the scaling of the corresponding Cramer-Rao bound ($\approx 1/\sqrt{2n}$ from Eq.~(\ref{cr}) can be achieved by such a scheme, albeit with a larger scaling factor.

However, it is not at all clear that on can similarly achieve a  $1/n$ scaling for {\it entangled} coherent states, despite the corresponding quantum Cramer-Rao bound having such a scaling  \cite{jooprl, hirota}.  This is essentially because the first part of the superposition in Eq.~(\ref{ecs}) does not see any phase shift, due to the vacuum contribution of the second mode.  This vacuum contribution adds significant phase noise.  For example, noting that the canonical phase distribution of a single mode field $|\psi\rangle$ is given by 
\[ p_C(\theta) = (1/2\pi) \left| \sum_m \langle m|\psi\rangle e^{-im\theta} \right|^2, \]
it is straightforward to calculate the joint canonical phase distribution $p_C(\theta_1,\theta_2)$ for the phase-shifted ECS $e^{-iN_2\phi}|\psi_\alpha\rangle$.  The  relative phase distribution $p_C(\phi_r)$, for $\phi_r:=\theta_2-\theta_1$, is then found by a suitable integration, as
\[  p_C(\phi_r) = \frac{1 + e^{-\alpha^2[1-\cos(\phi_r-\phi)]}\cos[\alpha^2\sin(\phi_r-\phi)]}{2\pi(1+e^{-\alpha^2})}
\]
For large $\alpha$ this is approximated in the neighbourhood of the phase shift $\phi$ by a mixture of a uniform distribution $1/2\pi$ with a Gaussian distribution of variance $1/\alpha^2=2/n$.  The presence of this uniform background distribution arises from the vacuum component of the ECS, and means that the above iterative approach cannot be straightforwardly applied in analogous manner directly analogous to the case of factorisable coherent states (where there was no such background term).  

Hence, unlike NOON states, it is perhaps not possible to achieve a $1/n$ scaling of phase resolution for entangled coherent states.  However, further investigation is required in this regard.

\subsection{Small average photon number}

The quantum Cramer-Rao bounds and information-theoretic bounds, for $\rmse_\phi$ and $\rmse$ respectively, are plotted in Fig.~3 for relatively small values of the total average photon number $n$, for both entangled coherent states and factorisable coherent states.  It is seen that the bound on the average performance for an ECS probe state (upper blue solid curve) always falls above that for the factorisable case.  Hence, as for the case of large $n$ discussed above, it appears that for ECS to gain the advantage suggested by the Cramer-Rao bounds (dotted curves), it is necessary to utilise them in more sophisticated phase estimation schemes (such as iterative schemes), using multicomponent probe states.

\section{Conclusions}

The main conclusion to be drawn from the above is that the quantum Cramer-Rao bound for $\rmse_\phi$ in Eq.~(\ref{cr}) only has direct operational significance for estimates that are locally unbiased in some neighbourhood of $\phi$.  Hence caution must be taken when using this bound, particularly if the range over which the estimate is unbiased is unknown.  Further illustration and discussion of this point is given elsewhere \cite{prx,berrypra}. 

The information-theoretic bounds for mutual information and $\rmse$ in Eqs.~(\ref{inf}) and (\ref{avbound}) are, in contrast, completely general.  They show that the relevant quantity for maximising average performance is not the variance of the generator, but its entropy (or, more generally, its $G$-asymmetry).  They may also be used to show, in some cases, that iterative implementations of proposed schemes can actually achieve the promise suggested by the corresponding Cramer-Rao bound --- but typically a larger scaling constant is required.  The information-theoretic bounds 
have been used elsewhere to find resolution bounds for general optical probe states and atomic qubit states, and to evaluate various nonlinear schemes proposed in the literature \cite{prx}.

Application of the new bounds to ECS probe states raises a question as to their average performance in comparison to unentangled coherent states.  As discussed in Sec.~V, a demonstration of superior performance will require more sophisticated estimation schemes than those based on using single-component probe states, and possibly even than those based on simple iterative implementations using multicomponent probe states.  This is an important challenge for future research in ECS-metrology.

Finally, it should be noted that the effects of noise and loss have not been considered here.  Entangled coherent states appear to have greater resilience than NOON states in this regard \cite{jooprl,joopra}, which may give them an advantage in realistic scenarios --- particularly if the challenge in the above paragraph can be met.

~\\

\acknowledgments  I thank H. Wiseman, A. Lund, D. Berry, and M. Zwierz for helpful discussions. This research was supported by the ARC Centre of Excellence CE110001027.

\end{document}